\documentclass[11pt,a4paper]{article}
\usepackage{jheppub}
\usepackage{amsmath,amssymb}
\usepackage{graphicx}
\usepackage{booktabs}
\usepackage{array}
\usepackage{bm}
\usepackage[lining]{OldStandard}
\newcommand{\Vna}{V_{n,\alpha}}
\newcommand{\Kn}{K_n}
\newcommand{\eps}{\varepsilon}
\newcommand{\ord}{\mathcal{O}}
\newcommand{\Nmu}{\mathcal{N}_{\mu}}

\newcommand{\slopeAgree}{\textbf{[agree]}}

\newcommand{\massA}{\textbf{[massA]}}

\newcommand{\massB}{\textbf{[massB]}}

\newcommand{\quarterA}{\textbf{[quarterA]}}
\newcommand{\quarterB}{\textbf{[quarterB]}}

\renewcommand{\slopeAgree}{$10^{-7}$ per unit length}

\renewcommand{\massA}{16.81}

\renewcommand{\massB}{13.47}

\renewcommand{\quarterA}{6.57}
\renewcommand{\quarterB}{10.29}

\begin{document}

\title{Quantum vacuum inversion and multi-kink unbinding in the non-degenerate double sine-Gordon model}

\author[a]{Jonathan Lozano-Mayo}
\affiliation[a]{Weinberg Institute for Theoretical Physics, Department of Physics, The University of Texas at Austin, Austin, Texas 78712, USA}
\emailAdd{jonathanloz@utexas.edu}

\abstract{Quantum fluctuations can unbind a composite soliton while the correction to its mass stays small. We show this beyond perturbation theory for the multi-kinks of the non-degenerate double sine-Gordon model. At one loop the vacuum at the centre of the multi-kink becomes the true vacuum at a coupling of order the classical lifting of the vacua. Tensor-network calculations of the lattice theory show that this inversion survives beyond one loop and has a continuum limit, and that the Gaussian effective potential captures it. On finite chains the multi-kink unbinds when the tension of the string of central vacuum between its two halves falls through zero.}


\maketitle

\section{Introduction}

Quantum fluctuations shift the mass of an elementary kink, and at weak coupling the shift is a small fraction of the classical mass for the $\phi^4$ and sine-Gordon kinks~\cite{DHN:1974,Coleman:1985,Rajaraman:1982,Manton:2004}. A multi-kink is held together differently. The multi-kink $K_n$ is the static solution of topological charge $Q=n$ that joins two global minima of the potential, the primary vacua, through $n$ sub-kinks. Between consecutive sub-kinks the field stays close to a local minimum that lies above the global ones. We call the vacuum built on it a lifted vacuum, and each such interval a string of that vacuum. The multi-kink exists because the tension of its strings balances the repulsion between its sub-kinks, and quantum fluctuations enter both sides of that balance. Whether it stays bound at all is then a quantum question.

The energy of a string is part of the mass of the soliton that carries it. The string has a classical tension, the classical energy density of the lifted vacuum above the primary one, and a quantum tension, the difference of the zero-point energy densities of the two vacua. The quantum tension is usually a small correction to the classical one, but not always. In the scheme in which the primary meson mass receives no one-loop correction, a vacuum whose meson mass differs from the primary one has a lower zero-point energy density than the primary vacuum. The quantum tension is then negative, and the one-loop energy of a soliton decreases as the region of that vacuum inside it grows~\cite{Weigel:2017b,Weigel:2018}. When the two vacua are degenerate nothing opposes the growth. The one-loop energy then has no lower bound at any coupling and the soliton is unstable, as found for the $\phi^6$ kink between inequivalent vacua~\cite{Weigel:2017b} and for kinks of the $\phi^8$ model~\cite{Takyi:2020}. Among the composite kinks of the Shifman--Voloshin model, which all have the same classical energy, quantum corrections also favour separated constituents~\cite{AlonsoIzquierdo:2004,AlonsoIzquierdo:2006}, whose one-loop energy falls without bound as the region of degenerate vacuum between them grows~\cite{Weigel:2018}.

When the vacuum inside the soliton is lifted, growing the interior has a classical cost. For a $\phi^6$ kink with a lifted interior minimum the one-loop correction to the mass is negative and grows only logarithmically in magnitude as the lifting is removed~\cite{AlonsoIzquierdo:2011,Graham:2022}. Its classical mass is finite, so as long as the lifting is nonzero the kink is stable at weak enough coupling~\cite{Graham:2022}, unlike the degenerate kink. The binding of such a soliton is decided by the total tension of its string, the classical cost of the lifted vacuum plus the negative quantum tension. At one loop the total tension changes sign at a finite coupling, while the correction to the mass stays bounded.

Where the total tension vanishes its classical and one-loop parts are equal and opposite, so the corrections beyond one loop decide on which side of zero it lies. We ask whether the sign change survives them. Our model is the non-degenerate double sine-Gordon model~\cite{LozanoMayo:2024dsg}. Its two parameters, $\alpha$ and $n$, set the lifting of the local minima and the ratio of the meson masses, and its multi-kinks carry several strings of lifted vacuum. The meson of the primary vacuum is light, and the quantum tension grows with the logarithm of the mass ratio, which is largest for the string at the centre. We put the normal-ordered Hamiltonian on a lattice and solve it with tensor networks, which do not rely on the loop expansion, and compare with one loop and with the Gaussian effective potential of the same Hamiltonian. We compute the vacuum energy densities on the infinite chain and follow the ground state of the four-kink sector on finite chains. The sign change survives: the vacua reorder at a coupling below the one-loop value by a relative amount of order $\alpha$, and the multi-kink unbinds there. The coupling at which the vacua reorder is extrapolated to the continuum for every $n$ and $\alpha$ studied, while the finite-chain calculations are for $n=4$ and $\alpha=0.3$ at two lattice spacings.

Section~\ref{sec:model} defines the model and its multi-kinks, section~\ref{sec:quantum} the quantum theory with its one-loop and Gaussian vacuum energy densities and the one-loop mass of the multi-kink, section~\ref{sec:vacua} the nonperturbative vacuum energy densities, and section~\ref{sec:string} the finite-chain ground states.

\section{The model and its multi-kinks}
\label{sec:model}

For a real scalar field $\phi$ with coupling $\beta$ we use the rescaled field $\chi=\beta\phi$ throughout, so that the classical energy of a static configuration is $E_{\rm cl}[\chi]/\beta^2$ with $E_{\rm cl}=\int dx[\frac12\chi'^2+\Vna(\chi)]$. The potential is
\begin{equation}
  \Vna(\chi)=\frac{(1-\cos\chi+\alpha)\,(1+\cos(\chi/n))}{2(1+\alpha/2)},\qquad n\ \text{even},
  \label{eq:V}
\end{equation}
with global minima at $\chi=\pm n\pi$ and, for $\alpha>0$, $n-1$ local minima near $\chi=2\pi i$ with $|i|<n/2$, lifted above them by an amount of order $\alpha$. We call the global minima the primary vacua and the local minima the lifted vacua, and keep the names for the corresponding quantum vacua. The name secondary vacuum is used in refs.~\cite{Weigel:2018,Takyi:2020} for a vacuum degenerate with the primary one. The true vacuum is the one of lowest energy density, which section~\ref{sec:quantum} decides.

The potential is $2n\pi$ periodic, non-negative, and vanishes only at the odd multiples of $n\pi$. Around $\chi=n\pi+\delta$ the factor $1+\cos(\chi/n)$ vanishes quadratically while $1-\cos\chi+\alpha$ tends to $\alpha$,
\begin{equation}
\begin{aligned}
  \Vna(n\pi+\delta)&=\frac{\alpha\,\delta^2}{2n^2(2+\alpha)}+\frac{6-\alpha/n^2}{24\,n^2(2+\alpha)}\,\delta^4+\ord(\delta^6),\\
  m_0^2&=\frac{\alpha}{n^2(2+\alpha)},
\end{aligned}
  \label{eq:m0}
\end{equation}
so at $\alpha=0$ the primary minimum is quartic, and the quadratic term that $\alpha$ creates is small for small $\alpha$ or large $n$. The lifted minima sit at $\chi_i=2\pi i+\ord(\alpha)$ with
\begin{equation}
\begin{aligned}
  \eps_i^{\rm cl}&=\Vna(\chi_i)=\alpha\,m_i^2+\ord(\alpha^2),\\
  m_i^2&=\Vna''(\chi_i)=\frac{1+\cos(2\pi i/n)}{2+\alpha}+\ord(\alpha),
\end{aligned}
  \label{eq:secondary}
\end{equation}
so to leading order in $\alpha$ the classical tension of a string of vacuum $i$ is $\alpha m_i^2/\beta^2$. We call $\eps_i^{\rm cl}$ the lifting of vacuum $i$. The lifted vacuum $i=0$ at $\chi=0$ is the central vacuum, and we write $m_{0,\rm lift}$ for its meson mass, since $m_0$ is that of the primary vacuum. Table~\ref{tab:betac} lists exact values.

\begin{figure}[t]
\centering
\includegraphics{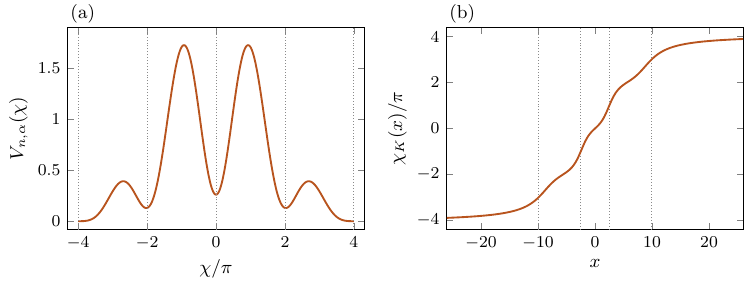}
\caption{The model at $n=4$, $\alpha=0.3$. (a) The potential of eq.~\eqref{eq:V}. Dotted lines mark the primary vacua at $\chi=\pm4\pi$ and the lifted ones at $0$ and $\pm2\pi$. (b) Classical profile of the multi-kink $K_4$, from eq.~\eqref{eq:quadrature}. Dotted lines mark the sub-kink centres, where $\chi_K$ crosses odd multiples of $\pi$.}
\label{fig:profile}
\end{figure}

The topological charge of a configuration is $Q=[\chi(+\infty)-\chi(-\infty)]/2\pi$, the net number of sub-kinks it contains. For $\Kn$, with $\chi(\pm\infty)=\pm n\pi$, the potential vanishes only at the end points, and the first integral $\chi'^2=2\Vna$ gives the profile $\chi_K(x)$ by quadrature,
\begin{equation}
  x(\chi)=\int_0^{\chi}\frac{d\chi'}{\sqrt{2\Vna(\chi')}},\qquad
  M_{\rm cl}=\int_{-n\pi}^{n\pi}\sqrt{2\Vna(\chi)}\,d\chi ,
  \label{eq:quadrature}
\end{equation}
and the field passes slowly through each lifted minimum, where $\sqrt{2\Vna}\simeq\sqrt{2\alpha}\,m_i$. Figure~\ref{fig:profile} shows $K_4$ for $\alpha=0.3$. The length $\ell_i$ over which the field dwells near $\chi_i$ follows from the linearization around $\chi_i$~\cite{LozanoMayo:2024dsg}. The two sub-kinks at the ends of the string both raise the field by $2\pi$, so they repel, with an interaction energy $A_ie^{-m_i\ell_i}$ in $E_{\rm cl}$. Balancing the force $m_iA_ie^{-m_i\ell_i}$ against the tension $\eps_i^{\rm cl}$ gives
\begin{equation}
  \ell_i=\frac{1}{m_i}\ln\frac{A_im_i}{\eps_i^{\rm cl}}.
  \label{eq:ellcl}
\end{equation}
For $n=4$ the sub-kink positions given by eq.~\eqref{eq:quadrature} follow eq.~\eqref{eq:ellcl} with a constant $\ln(A_0m_{0,\rm lift})\simeq3.4$ over $0.01\le\alpha\le0.5$, while $\eps_0^{\rm cl}$ changes by a factor of $40$. The tails of $\Kn$ approach $\pm n\pi$ as $e^{-m_0|x|}$ over the much longer distance $1/m_0=n\sqrt{(2+\alpha)/\alpha}$, which is $11$ for $n=4$ and $\alpha=0.3$.

The double sine-Gordon model, a sine-Gordon potential with a second cosine of commensurate frequency, is well studied in its own right. Form-factor perturbation theory and the truncated conformal space approach give its spectrum and phase transitions~\cite{Delfino:1998,Bajnok:2001,Takacs:2006} and those of its multi-frequency generalizations~\cite{Toth:2004}. Semiclassical form factors of its kinks give the masses of the mesons they bind into~\cite{Mussardo:2004}, and its classical kink dynamics is known in detail~\cite{Campbell:1986}. A lifted vacuum in such a model decays~\cite{Coleman:1977,Mussardo:2004}. The potential of eq.~\eqref{eq:V} differs in that a single soliton carries $n-1$ strings of lifted vacuum whose classical tension is small by choice, while the meson of the primary vacuum is anomalously light.

\section{One-loop and Gaussian energies}
\label{sec:quantum}

\subsection{Hamiltonian and scheme}

The quantum model is
\begin{equation}
  H=\int dx\left[\tfrac12\pi^2+\tfrac12(\partial_x\phi)^2+\frac{1}{\beta^2}\,\Nmu\!\left[\Vna(\beta\phi)\right]\right],\qquad
  [\phi(x),\pi(y)]=i\delta(x-y),
  \label{eq:H}
\end{equation}
where $\Nmu$ denotes normal ordering with respect to the free field of mass $\mu$. In terms of $\chi=\beta\phi$ and $p=\beta\pi$ the commutator is $[\chi,p]=i\beta^2$ and $H=\beta^{-2}\int[\frac12p^2+\frac12\chi'^2+\Nmu\Vna(\chi)]$, so $\beta^2$ is the loop-counting parameter and classical energies are $\ord(\beta^{-2})$. Normal ordering removes the only ultraviolet divergence of the theory. Since $\Vna$ is a sum of four cosines,
\begin{equation}
\begin{gathered}
  \Vna(\chi)=\sum_k c_k\cos(k\chi),\\
  k\in\{0,\,1,\,\tfrac1n,\,1+\tfrac1n,\,1-\tfrac1n\},\\
  c_0=c_{1/n}=\frac{1+\alpha}{2+\alpha},\quad c_1=-\frac{1}{2+\alpha},\\
  c_{1\pm1/n}=-\frac{1}{2(2+\alpha)},
\end{gathered}
  \label{eq:cosines}
\end{equation}
and $\Nmu[\cos k\chi]=\cos(k\chi)\exp[k^2\beta^2G_\mu/2]$ with $G_\mu=\langle\phi^2\rangle_\mu$ the regularized free propagator at coincident points, the scheme multiplies each cosine by a different factor. The prescriptions at $\mu$ and $\mu'$ are related by
\begin{equation}
  \Nmu[\cos k\chi]=\left(\frac{\mu'}{\mu}\right)^{k^2\beta^2/4\pi}\mathcal{N}_{\mu'}[\cos k\chi],
  \label{eq:muscale}
\end{equation}
since $G_\mu-G_{\mu'}=(1/2\pi)\ln(\mu'/\mu)$. The Hamiltonian written at $\mu'$ with the same coefficients $c_k$ is therefore a different theory. To keep the theory fixed, each coefficient must be rescaled by its own power of $\mu'/\mu$, and such a change of the four coefficients is not a change of $\alpha$ or $\beta$. The normal-ordering prescription is part of the definition of the quantum theory, as it is for every multi-frequency sine-Gordon model~\cite{Delfino:1998,Bajnok:2001}. The theory studied in this paper is $\mathcal{N}_{m_0}\Vna$, the coefficients $c_k$ of eq.~\eqref{eq:cosines} normal ordered at the classical mass of the primary meson. That choice is the no-tadpole scheme of the vacuum-polarization literature~\cite{Graham:2009,Weigel:2017}. At the primary vacuum normal ordering at $m_0$ cancels the one-loop tadpole in the meson self-energy. Since $\Vna$ is even about $n\pi$ there is no cubic vertex either, so the primary meson mass remains $m_0$ through one loop. The parameter $\alpha$ sets the classical lifting of the local minima. With the same coefficients normal ordered at the central meson mass $m_{0,\rm lift}$ instead, the one-loop correction vanishes at the central vacuum and, for $m_{0,\rm lift}\gg m_0$, lowers the energy density of the primary vacuum by about $m_{0,\rm lift}^2/8\pi$, more than that of any lifted vacuum, so no inversion occurs at one loop.

\subsection{One-loop inversion}
\label{sec:oneloop}

For a homogeneous field $\chi$ with $m^2(\chi)=\Vna''(\chi)>0$, the one-loop energy density in the scheme $\Nmu$ is (appendix~\ref{app:oneloop})
\begin{equation}
  \eps^{(1)}(\chi)=\frac{1}{8\pi}\left[m^2(\chi)-\mu^2-m^2(\chi)\ln\frac{m^2(\chi)}{\mu^2}\right].
  \label{eq:eps1}
\end{equation}
The zero-point energy density of a free field is a concave function of its mass squared, and normal ordering at $\mu$ subtracts the tangent to that function at $\mu^2$. The bracket in eq.~\eqref{eq:eps1} is the difference between the function and its tangent, so $\eps^{(1)}\le0$ with equality only at $m^2=\mu^2$, and the energy density of a vacuum whose meson is much heavier than $\mu$ is lowered by about $m^2\ln(m^2/\mu^2)/8\pi$. With $\mu=m_0$ the energy density of the primary vacuum is unshifted. The energy density of the lifted vacuum $i$ minus that of the primary vacuum is
\begin{equation}
  \Delta\eps_i(\beta^2)=\frac{\eps_i^{\rm cl}}{\beta^2}+\frac{m_0^2}{8\pi}\left[x_i-1-x_i\ln x_i\right],\qquad x_i=\frac{m_i^2}{m_0^2}.
  \label{eq:gap}
\end{equation}
The first term is the classical tension of a string of vacuum $i$, the second its quantum tension, and the ordering of the two vacua inverts where the two cancel,
\begin{equation}
\begin{aligned}
  \beta_{c,i}^2&=\frac{8\pi\,\eps_i^{\rm cl}}{m_0^2\,[x_i\ln x_i-x_i+1]}
  \simeq\frac{8\pi\alpha}{\ln x_i-1},\\
  x_i&\simeq\frac{n^2(1+\cos 2\pi i/n)}{\alpha}.
\end{aligned}
  \label{eq:betac}
\end{equation}
The second form uses eq.~\eqref{eq:secondary}. The central vacuum $i=0$ has the largest mass ratio and inverts first, at
\begin{equation}
  \beta_c^2\simeq\frac{8\pi\alpha}{\ln(2n^2/\alpha)-1},
  \label{eq:betac-intro}
\end{equation}
a coupling of order $\alpha$ rather than of order the sine-Gordon scale $8\pi$, because the classical tension is proportional to $\alpha$ while the quantum tension depends on $\alpha$ only through $\ln x_i$. We call $\beta_c^2(\alpha)$ the inversion line. Beyond it the central vacuum has the lower energy density, and the runaway of the degenerate case becomes an inversion of the ordering of the two vacua at a finite, small coupling. The dependence on $\alpha$ is close to linear because the logarithm varies slowly, and $n$ enters only through $\ln n^2$, so for fixed $\alpha$ the inversion line moves to smaller coupling as $n$ grows (table~\ref{tab:betac_all}). Nothing in eq.~\eqref{eq:betac} uses the form of $\Vna$ beyond the lifting and the meson masses of the two vacua.

\begin{table}[t]
\centering
\caption{Classical data of $\Vna$ from eqs.~\eqref{eq:m0}--\eqref{eq:quadrature}, with $x_0=m_{0,\rm lift}^2/m_0^2$ the squared meson-mass ratio of the central vacuum to the primary one. Last column: one-loop correction to the mass of $\Kn$ relative to the classical mass at $\beta^2=1$ (appendix~\ref{app:vpe}). The inversion couplings are collected in table~\ref{tab:betac_all}.}
\label{tab:betac}
\small
\begin{tabular}{cccccccc}
\toprule
$n$ & $\alpha$ & $m_0^2$ & $\eps_0^{\rm cl}$ & $m_{0,\rm lift}^2$ & $x_0$ & $M_{\rm cl}$ & $E^{(1)}/M_{\rm cl}$\\
\midrule
4 & 0.05 & 0.00152 & 0.04878 & 0.9741 & 639 & 21.267 & $-0.085$\\
4 & 0.1  & 0.00298 & 0.09524 & 0.9494 & 319 & 21.752 & $-0.059$\\
4 & 0.2  & 0.00568 & 0.18182 & 0.9034 & 159 & 22.530 & $-0.035$\\
4 & 0.3  & 0.00815 & 0.26087 & 0.8614 & 105.7 & 23.158 & $-0.026$\\
4 & 0.5  & 0.01250 & 0.40000 & 0.7875 & 63 & 24.149 & $-0.017$\\
8 & 0.1  & 0.00074 & 0.09524 & 0.9516 & 1279 & 43.210 & $-0.070$\\
8 & 0.3  & 0.00204 & 0.26087 & 0.8675 & 425.7 & 46.071 & $-0.033$\\
\bottomrule
\end{tabular}
\end{table}

\subsection{Beyond one loop: the Gaussian effective potential}
\label{sec:gep}

The logarithm $\ln x_i$ that sets the one-loop tension also controls the next orders. The variational estimate that resums the tadpole corrections of the cosine operators is the Gaussian effective potential (GEP)~\cite{Stevenson:1985}. For a Gaussian trial state of mass $\Omega$ centred at $\bar\chi$ the expectation value of the energy density of eq.~\eqref{eq:H} is
\begin{equation}
  \eps_G(\bar\chi,\Omega)=\frac{\Omega^2-\mu^2}{8\pi}+\frac{1}{\beta^2}\sum_k c_k\left(\frac{\Omega}{\mu}\right)^{k^2\beta^2/4\pi}\cos(k\bar\chi),
  \label{eq:gep}
\end{equation}
minimized over $\Omega$ (appendix~\ref{app:oneloop}). The Gaussian average of each cosine carries the factor $e^{y_k}$ with
\begin{equation}
  y_k=\frac{k^2\beta^2}{8\pi}\ln\frac{\Omega^2}{\mu^2},
  \label{eq:yk}
\end{equation}
and one loop is the linearization of eq.~\eqref{eq:gep} at $\Omega^2=m^2(\bar\chi)$, which keeps $1+y_k$. At the primary vacuum the gap equation is solved by $\Omega=\mu=m_0$ for all $\beta^2$, every $y_k$ vanishes, and $\eps_G=0$. The tadpole that would stiffen the quartic minimum has been absorbed into the couplings by the scheme. At a lifted vacuum $\Omega^2$ is of order $m_i^2$ and $y_k\simeq k^2\beta^2\ln x_i/8\pi$, which at the one-loop inversion line becomes, by eq.~\eqref{eq:betac},
\begin{equation}
  y_k\big|_{\beta^2=\beta_c^2}=k^2\alpha\,\frac{\ln x_i}{\ln x_i-1}\simeq k^2\alpha .
  \label{eq:ykline}
\end{equation}
The expansion at the inversion line is an expansion in $\alpha$, so for $\alpha\ll1$ the one-loop coupling is correct to relative order $\alpha$. The sign of the correction is fixed. The factor $e^{y_k}$ exceeds its linearization $1+y_k$, and at a lifted vacuum the cosines with the largest $k$, which gain most, carry negative coefficients. The Gaussian energy density of a lifted vacuum therefore lies below its one-loop value even at $\Omega=m$, and the minimization over $\Omega$ lowers it further. The coupling at which the two energy densities cross therefore moves to smaller values, by $4$ percent at $\alpha=0.05$ and $22$ percent at $\alpha=0.3$ for $n=4$ (table~\ref{tab:betac_all}).

\subsection{One-loop mass of the multi-kink}
\label{sec:mass}

The one-loop correction $E^{(1)}$ to the mass of $\Kn$ is the vacuum polarization energy of the fluctuation operator around the classical profile, which we evaluate by the spectral method (appendix~\ref{app:vpe}). It is negative and between $1.7$ and $8.5$ percent of $M_{\rm cl}$ at $\beta^2=1$ (table~\ref{tab:betac}). Much of it comes from the strings. Since $\eps^{(1)}$ vanishes at the primary vacuum, a string of vacuum $i$ contributes about $\ell_i\,\eps^{(1)}(\chi_i)$, its classical length times the quantum tension of eq.~\eqref{eq:gap}. The binding of the multi-kink depends on the same tension through the energy cost of lengthening a string, which is the total tension. The mass can therefore stay close to its classical value while the total tension changes sign.

\section{The inversion beyond perturbation theory}
\label{sec:vacua}

Neither approximation is controlled at $\alpha$ of order a few tenths, which is the range in which the sub-kink structure of $\Kn$ is resolved. We compute the vacuum energy densities nonperturbatively. The theory is put on a chain of spacing $a$ with the normal-ordering factors $Z_k=\exp[k^2\beta^2G_a(\mu)/2]$, where $G_a$ is the lattice propagator at coincident points. The lattice Hamiltonian is then the transcription of eq.~\eqref{eq:H} in the scheme $\Nmu$ with no free parameter, and as $a\to0$ the $Z_k$ reproduce the multiplicative renormalization of the sine-Gordon operators~\cite{Coleman:1975}. The one-loop and Gaussian energy densities are evaluated on the same lattice with the same $Z_k$, so the three calculations refer to one Hamiltonian.

The energy density of a homogeneous vacuum is obtained on the infinite chain with the variational uniform matrix product state (VUMPS) algorithm~\cite{ZaunerStauber:2018} (appendix~\ref{app:lattice}), optimized in the basin of $\chi=n\pi$ or of $\chi=0$. A translation-invariant state leaves its basin only by moving every site at once, so the optimization returns the energy density of the metastable vacuum when it is not the lowest. We write $\eps(\chi)$ for the energy density of the vacuum at $\chi$, so that $\Delta\eps_0=\eps(0)-\eps(n\pi)$.

\begin{figure}[t]
\centering
\includegraphics[width=\textwidth]{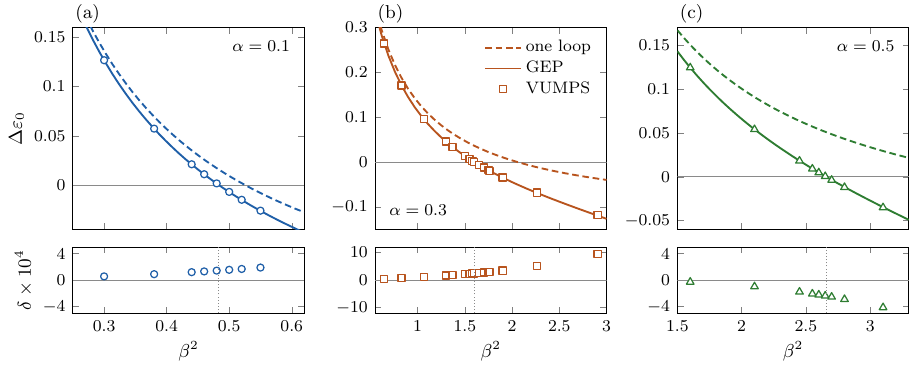}
\caption{Energy-density difference $\Delta\eps_0=\eps(0)-\eps(4\pi)$ of the central and primary vacua for $n=4$ on the lattice with $a=1/2$, per unit length in the units of the classical model. Top: VUMPS values (open symbols), the lattice Gaussian effective potential (solid) and lattice one loop (dashed), all in the scheme $\mu=m_0$. Bottom: $\delta=\Delta\eps_0^{\rm VUMPS}-\Delta\eps_0^{\rm GEP}$, and the dotted line marks the GEP crossing.}
\label{fig:vacua}
\end{figure}

Figure~\ref{fig:vacua} shows $\Delta\eps_0$ for $n=4$ at $a=1/2$ and three values of $\alpha$. The nonperturbative values change sign in every case, at the couplings of table~\ref{tab:betac_all}, and the Gaussian effective potential follows them over the whole range. Its crossing is the difference of two variational energies and has no error estimate of its own, so the agreement is a result of the nonperturbative calculation. The residual $\delta$, the difference between the non-Gaussian corrections to the energy densities of the two vacua, is a few times $10^{-4}$ near the crossings.

\begin{table}[t]
\centering
\caption{Inversion coupling $\beta_c^2$ of the central vacuum by method and regularization. The continuum values at one loop and from the Gaussian effective potential (GEP) are eqs.~\eqref{eq:betac} and \eqref{eq:gep} with the classical data of table~\ref{tab:betac}. The values at lattice spacing $a$ are the same expressions with the lattice propagator of eq.~\eqref{eq:Ga}, and the VUMPS values are the crossings of the nonperturbative vacuum energy densities. The last row is the continuum limit of the VUMPS crossing described in the text, with the uncertainty in the last digit in brackets.}
\label{tab:betac_all}
\small
\setlength{\tabcolsep}{2.6pt}
\begin{tabular}{llccccccc}
\toprule
 & & \multicolumn{5}{c}{$n=4$, $\alpha=$} & \multicolumn{2}{c}{$n=8$, $\alpha=$}\\
\cmidrule(lr){3-7}\cmidrule(lr){8-9}
method & regularization & $0.05$ & $0.1$ & $0.2$ & $0.3$ & $0.5$ & $0.1$ & $0.3$\\
\midrule
one loop & continuum & 0.2305 & 0.5287 & 1.2412 & 2.0740 & 4.0411 & 0.4087 & 1.4948\\
GEP & continuum & 0.2213 & 0.4867 & 1.0498 & 1.6145 & 2.6845 & 0.3819 & 1.2260\\
one loop & $a=1/2$ & 0.2292 & 0.5255 & 1.2327 & 2.0591 & 4.0101 & 0.4067 & 1.4867\\
GEP & $a=1/2$ & 0.2200 & 0.4835 & 1.0416 & 1.6003 & 2.6568 & 0.3800 & 1.2181\\
VUMPS & $a=1/2$ & 0.2201 & 0.4839 & 1.0429 & 1.6022 & 2.6540 & 0.3800 & 1.2177\\
GEP & $a=1/4$ & 0.2209 & 0.4856 & 1.0469 & 1.6095 & 2.6746 & 0.3813 & 1.2232\\
VUMPS & $a=1/4$ & 0.2209 & 0.4859 & 1.0483 & 1.6114 & 2.6705 & 0.3812 & 1.2228\\
GEP & $a=1/8$ & 0.2211 & 0.4863 & 1.0489 & 1.6129 & 2.6813 & 0.3817 & 1.2251\\
VUMPS & $a=1/8$ & 0.2211 & 0.4866 & 1.0502 & 1.6149 & 2.6774 & 0.3812 & 1.2245\\
VUMPS & $a\to0$ & 0.2213(1) & 0.4870(1) & 1.0511(2) & 1.6164(2) & 2.6804(4) & 0.3815(4) & 1.2254(2)\\
\bottomrule
\end{tabular}
\end{table}

The lattice spacing enters the one-loop and Gaussian energy densities through $G_a$ alone. Repeating the nonperturbative calculation at $a=1/4$ and $1/8$ moves its crossings together with the Gaussian ones, by the change of $G_a(\mu)$. Their difference $\delta_c(a)=\beta_{c,\rm V}^2-\beta_{c,\rm G}^2$, with $\beta_{c,\rm G}^2$ the Gaussian crossing, lies between $-0.0041$ and $+0.0020$ in every column of table~\ref{tab:betac_all}. The continuum Gaussian crossings are known from eq.~\eqref{eq:gep}, so the continuum limit is taken on $\delta_c$ rather than on $\beta_c^2$ itself. The lattice propagator expands as $G_a(m)=(1/2\pi)\ln(C/am)+\ord(a^2m^2\ln am)$ with a constant $C$, so a bulk energy density approaches its continuum value with corrections of order $a^2\ln a$ and $a^2$. At $\alpha=0.3$, fitting $c_0+c_1a^2+c_2a^2\ln a$ to the Gaussian crossings at the three lattice spacings gives the continuum value $c_0=1.61449$, against the exact $1.61446$. Between $a=1/4$ and $1/8$ the remainder $\delta_c$ changes by at most $4\times10^{-4}$, within the bond-dimension uncertainty of the crossings at $a=1/8$ (appendix~\ref{app:lattice}). Extrapolating it with either correction term, or with none, gives the continuum crossings of table~\ref{tab:betac_all}, whose uncertainty combines the spread of the three extrapolations with that error.

For $n=4$ the relative excess of the one-loop crossing $\beta^2_{c,\rm 1L}$ over the nonperturbative one $\beta^2_{c,\rm V}$, $(\beta^2_{c,\rm 1L}/\beta^2_{c,\rm V}-1)/\alpha$, rises from $0.83$ at $\alpha=0.05$ to $1.02$ at $\alpha=0.5$ (table~\ref{tab:betac_all}), the order-$\alpha$ error that eq.~\eqref{eq:ykline} predicts. The one-loop inversion line is the $\alpha\to0$ limit of the nonperturbative one, and over this range their ratio is close to $1+\alpha$. The Gaussian estimate also gives the vacua at $\chi=\pm2\pi$ a lower energy density than the one at $\chi=4\pi$ for $\beta^2>2.14$ at $\alpha=0.3$, so beyond that coupling the primary vacuum is metastable against all three secondary vacua.

\section{The multi-kink as a bound state}
\label{sec:string}

\subsection{The central string as a collective coordinate}
\label{sec:beyond}

Below the inversion line the multi-kink is a bound state of its two halves, each a wall that carries half of the topological charge and joins the primary vacuum to the central one. When the two walls are a distance $\ell_0$ apart, large compared with $1/m_c$, the energy of the configuration is the sum of the masses of the two walls, $E_0$, the energy $\Delta\eps_0\,\ell_0$ of the string of central vacuum between them, and their interaction, which comes from the overlap of their tails. Here $m_c$ is the meson mass of the central vacuum, $m_{0,\rm lift}$ classically. Between the walls, at $X_L<x<X_R$ with $\ell_0=X_R-X_L$, the field is $\chi_c+\eta_L+\eta_R$. The tails are $\eta_L=-b_Le^{-m_c(x-X_L)}$ and $\eta_R=b_Re^{m_c(x-X_R)}$ with $b_L,b_R>0$, since the field approaches $\chi_c$ from smaller values and leaves it towards larger ones. In this overlap approximation the quadratic cross term of the tails in the energy density, $\eta_L'\eta_R'+m_c^2\eta_L\eta_R$, vanishes, so the leading interaction is read instead from the force between the walls, the stress $T_{11}=[\tfrac12\chi'^2-\Vna(\chi)]/\beta^2$ at a point between them. To quadratic order in the tails their squares cancel in it, and the cross term is
\begin{equation}
  \delta T_{11}=\frac{\eta_L'\eta_R'-m_c^2\eta_L\eta_R}{\beta^2}=\frac{2m_c^2b_Lb_R}{\beta^2}\,e^{-m_c\ell_0},
  \label{eq:dT11}
\end{equation}
independent of $x$. The force on the right wall is the interior stress minus the exterior one, $F=-\Delta\eps_0+\delta T_{11}$, and $F=-dE/d\ell_0$ integrates to
\begin{equation}
  E(\ell_0)=E_0+\Delta\eps_0\,\ell_0+A\,e^{-m_c\ell_0},\qquad A=\frac{2m_cb_Lb_R}{\beta^2}>0.
  \label{eq:Eell}
\end{equation}
The interaction is repulsive because both walls raise the field. Classically $\Delta\eps_0=\eps_0^{\rm cl}/\beta^2$ and $A=A_0/\beta^2$, and the minimum of $E$ is the string length of eq.~\eqref{eq:ellcl}. In the quantum theory we take eq.~\eqref{eq:Eell} as an effective description at large separation, valid as long as the separation changes slowly on the scale of the meson masses, and section~\ref{sec:charge4} tests it. In it $\Delta\eps_0$ is the difference of vacuum energy densities of section~\ref{sec:vacua} and $m_c$ the mass gap of the central vacuum, both properties of the homogeneous vacua.

In the quantum theory $\ell_0$ is a collective coordinate with a conjugate momentum $p_\ell$, and, treating each wall as rigid with half of the classical mass, the reduced mass is $\mu_r=M_{\rm cl}/(4\beta^2)$. Its Hamiltonian is
\begin{equation}
  H_{\rm coll}=\frac{p_\ell^2}{2\mu_r}+E(\ell_0),\qquad [\ell_0,p_\ell]=i,
  \label{eq:Hcoll}
\end{equation}%
and its ground state has a time-independent mean $\langle\ell_0\rangle$ and a finite spread.

Because $\Delta\eps_0$ vanishes at $\beta_c^2$, the string length departs from its classical value by much more than the $\ord(\beta^2)$ that a soliton coordinate usually does. The minimum of $E$ moves out as $\ln(1/\Delta\eps_0)$. Near the inversion line the potential rises linearly with slope $\Delta\eps_0$ from a steep wall at $\ell_w\simeq\ln(A/E_{\rm zp})/m_c$, with $E_{\rm zp}$ the zero-point energy of the coordinate. In this linear-confinement approximation the ground-state wave function is an Airy function, with
\begin{equation}
\begin{aligned}
  \langle\ell_0\rangle&\simeq\ell_w+1.56\,\Big(\frac{2\beta^2}{M_{\rm cl}\,\Delta\eps_0}\Big)^{1/3},\\
  E_{\rm zp}&\simeq2.34\,\Big(\frac{2\beta^2\Delta\eps_0^2}{M_{\rm cl}}\Big)^{1/3}.
\end{aligned}
  \label{eq:airy}
\end{equation}
The mean string length diverges at the inversion line as $\Delta\eps_0^{-1/3}$, and so does its spread, as for any pair confined by a linear potential whose tension is taken to zero. The description holds as long as $E_{\rm zp}$ is small compared with the meson masses of both vacua, that is, $(2\beta^2\Delta\eps_0^2/M_{\rm cl})^{1/3}\ll m_0$. Close to the inversion line it is, since at $\alpha=0.3$ and $\beta^2=1.58$ the nonperturbative tension $\Delta\eps_0=2.9\times10^{-3}$ gives $E_{\rm zp}\approx0.02$, against $m_0=0.09$.

The tension is the quantity in which the one-loop and nonperturbative vacuum energy densities differ most, and the mean string length is the observable most sensitive to it. Beyond the inversion line the slope changes sign and the energy of the sector decreases without bound as the string stretches. In a box with the field fixed at $\pm n\pi$ at the ends, in the ground state the two halves sit near the ends with the central vacuum between them. The halves survive as walls between the two vacua, and only their binding is lost. 

\subsection{Ground states of the $Q=4$ sector}
\label{sec:charge4}

The energy and the profile of the ground state of the $Q=4$ sector were computed at $\alpha=0.3$ and $a=1/2$ on a chain of $L=80$ with the field fixed at $\mp4\pi$ at the ends, for four couplings across the inversion line. The method is single-site density-matrix renormalization group (DMRG)~\cite{White:1992,Schollwock:2011} with subspace expansion (appendix~\ref{app:lattice}). The optimization is started from the classical multi-kink and from its two halves moved to the ends of the chain. The string lengths are measured as $\ell_i=a\sum_j\langle\Theta(\pi-|\chi_j-2\pi i|)\rangle$, with $\chi_j$ the field at site $j$ and $\Theta$ the step function: the length over which the field lies within $\pi$ of $2\pi i$.

\begin{table}[t]
\centering
\caption{$Q=4$ ground states at $\alpha=0.3$, $a=1/2$, $L=80$. $E$ is the energy in the units of the classical model, $\ell_0$ the length of the central string, $\ell_{\pm1}$ the common length of the two outer strings, which agree to within $4\times10^{-4}$ in these runs, and $M$ the mass of the bound multi-kink (section~\ref{sec:charge4}). The classical values are $\ell_0=5.00$, $\ell_{\pm1}=7.39$ and $M_{\rm cl}/\beta^2=17.81$ and $14.66$.}
\label{tab:charge4}
\begin{tabular}{lccccc}
\toprule
$\beta^2$ & $E$ & ground state & $\ell_0$ & $\ell_{\pm1}$ & $M$\\
\midrule
$1.30$ & $220.784$ & bound & $6.88$ & $9.23$ & $\massA$\\
$1.58$ & $217.430$ & bound & $10.93$ & $10.80$ & $\massB$\\
$1.75$ & $215.374$ & stretched & $43.7$ & $9.65$ & --\\
$1.90$ & $213.591$ & stretched & $49.4$ & $8.45$ & --\\
\bottomrule
\end{tabular}
\end{table}

Table~\ref{tab:charge4} gives the result. Both starting configurations converge to the same state, which is bound below the inversion line and stretched above it, with a central string that then covers most of the box. The multi-kink unbinds between $\beta^2=1.58$ and $1.75$, the interval that contains the nonperturbative crossing $1.6022$ and excludes the one-loop crossing $2.06$, which would keep it bound through $\beta^2=1.90$.

\begin{figure}[t]
\centering
\includegraphics{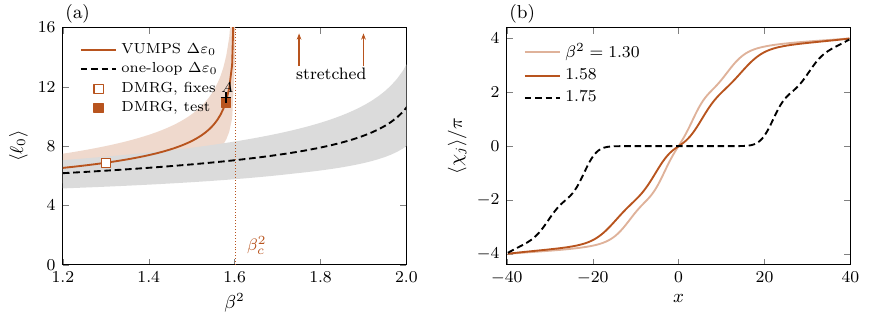}
\caption{$Q=4$ sector at $\alpha=0.3$ and $a=1/2$. (a) Length of the central string against $\beta^2$. Lines are the ground-state mean $\langle\ell_0\rangle$ of the collective Hamiltonian~\eqref{eq:Hcoll} with the potential of eq.~\eqref{eq:Eell}, using the tension from the VUMPS vacuum energy densities (solid) and from lattice one loop (dashed), with $m_c=1.16$ and $A$ fixed by the DMRG value at $\beta^2=1.30$ (open square). Each band is the envelope of $\langle\ell_0\rangle$ plus and minus the width of the ground-state wave function for $m_c=0.93$ and $1.16$. The filled square is the DMRG value at $\beta^2=1.58$, the cross is the result at $L=120$. At $\beta^2=1.75$ and $1.90$ the ground state is stretched and the string fills the box (arrows). The dotted line is the nonperturbative crossing. (b) Ground-state profiles $\langle\chi_j\rangle$ at $L=80$.}
\label{fig:charge4}
\end{figure}

The string lengths of the two bound states test eq.~\eqref{eq:Eell} with the two tensions (figure~\ref{fig:charge4}a). The repulsion strength $A$ is the one quantity in the model that the vacuum calculation does not supply. Its classical value, $\ln(A_0m_{0,\rm lift})\simeq3.4$ from section~\ref{sec:model}, with the classical $m_c$ and the nonperturbative tension gives $\langle\ell_0\rangle=7.2$ at $\beta^2=1.30$ and $11.7$ at $1.58$, five and eight percent above the measured values. We fix $A$ instead by the string length at $\beta^2=1.30$, with $m_c=1.16$ from the correlation length of the central vacuum or the classical $0.93$. With the classical $m_c$ this gives $A\simeq A_0/1.4$. At $\beta^2=1.58$ the model then predicts $\langle\ell_0\rangle=11.0$ to $11.4$ with the nonperturbative tension and $7.0$ with the one-loop tension, against $\ell_0=10.93$ at $L=80$ and $11.28$ at $L=120$ in the DMRG ground state (figure~\ref{fig:charge4}a).

Between the two bound states the outer strings grow as well (table~\ref{tab:charge4}), since their tensions $\Delta\eps_{\pm1}$ fall with $\beta^2$ in the same way. In the profiles of figure~\ref{fig:charge4}b, at the two lower couplings the multi-kink sits in the middle of the box with its four sub-kinks resolved and the central step widening.

The same chains give the mass of the multi-kink, as the energy of the $Q=4$ ground state minus that of the chain with both ends held at $+4\pi$. The difference removes the bulk energy and the boundary terms common to the two. The remaining finite-box contribution is the interaction of the tails with the ends of the chain, bounded by the $0.006$ change of $E(L)-L\eps(4\pi)$ between $L=80$ and $120$ in section~\ref{sec:finiteL}. The result (table~\ref{tab:charge4}) lies $5.6$ and $8.1$ percent below the classical $M_{\rm cl}/\beta^2$ at $\beta^2=1.30$ and $1.58$, against $3.4$ and $4.1$ percent at one loop. The correction is $1.6$ to $2$ times the one-loop estimate and still under a tenth of the mass where the central string has more than doubled its classical length.

\subsection{The box-length dependence}
\label{sec:finiteL}

\begin{figure}[t]
\centering
\includegraphics{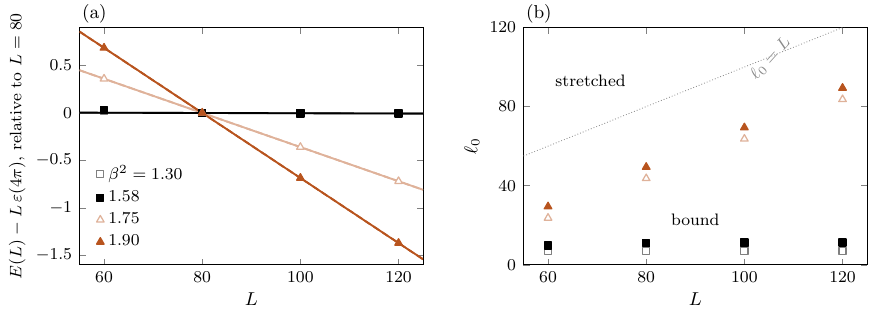}
\caption{Box-length dependence of the $Q=4$ ground state at $\alpha=0.3$, $a=1/2$. (a) $E(L)-L\eps(4\pi)$, with $\eps(4\pi)$ the infinite-chain energy density of the primary vacuum, for the four couplings. Solid lines are the linear fits of table~\ref{tab:finiteL}, and the dashed lines, hidden under them above the inversion line, have the slopes $\Delta\eps_0=\eps(0)-\eps(4\pi)$ from the infinite-chain vacuum energy densities. (b) Length of the central string against $L$.}
\label{fig:finiteL}
\end{figure}

The string length depends on how it is defined. The tension can be obtained without any such definition, from the dependence of the energy on the length of the box. In a static eigenstate momentum conservation makes $\langle T_{11}\rangle$ uniform, and in a box with the field fixed at the ends it equals $-dE/dL$, by the Hellmann--Feynman theorem for a uniform rescaling of the box. For the bound multi-kink $\langle T_{11}\rangle$ is the pressure $-\eps(n\pi)$ of the primary vacuum, so $dE/dL=\eps(n\pi)$, the quantum form of the classical scaling relation $\int T_{11}\,dx=0$~\cite{LozanoMayo:2026virial}. For the stretched state the same argument gives $dE/dL=\eps(0)$. The energy $E(L)-L\eps(4\pi)$, with that of a box filled with the primary vacuum subtracted, is therefore flat in $L$ for a bound multi-kink and has slope $\Delta\eps_0$ for two walls confining a slab of the true vacuum. Figure~\ref{fig:finiteL} shows it at $L=60$ to $120$ for the four couplings, together with the length of the central string, and table~\ref{tab:finiteL} gives the slopes.

Below the inversion line the subtracted energy is the mass of the multi-kink plus boundary terms that do not depend on $L$, and it stops changing once the box is large enough to hold the state. Between $L=80$ and $120$ it changes by less than $0.01$, against the $0.1$ to $2$ that a string filling the added length would cost. A box of $L=60$ compresses the state at $\beta^2=1.58$, where the multi-kink and its tails span some $60$ units, so the slopes below the inversion line are fitted to the three largest boxes. Above the inversion line the subtracted energy decreases linearly, with slopes equal to $\Delta\eps_0$ from the infinite-chain vacuum energy densities to within \slopeAgree. Repeating the test at $a=1/4$ with $L=80$ and $120$ gives the lower block of table~\ref{tab:finiteL}: the same two regimes, with the slopes above the inversion line again equal to $\Delta\eps_0$ of the $a=1/4$ chain to within $10^{-7}$. The bound strings are shorter at this spacing, $\ell_0=\quarterA$ and $\quarterB$ at $L=80$ against $6.88$ and $10.93$ at $a=1/2$, most at $\beta^2=1.58$, where the crossing has moved from $1.6022$ to $1.6114$ and the tension has grown from $2.91\times10^{-3}$ to $4.06\times10^{-3}$ (table~\ref{tab:finiteL}).

\begin{table}[t]
\centering
\caption{Box-length dependence of the $Q=4$ ground state at $\alpha=0.3$: slope of $E(L)-L\eps(4\pi)$, the difference $\Delta\eps_0=\eps(0)-\eps(4\pi)$ of the vacuum energy densities on the infinite chain at the same spacing, $d\ell_0/dL$, and the state. At $a=1/2$ the slopes are fitted, the stretched states over $L=60$ to $120$ and the bound ones over $L=80$ to $120$, and given with their standard errors, which for the stretched states are below $10^{-7}$. At $a=1/4$ they are taken between $L=80$ and $120$.}
\label{tab:finiteL}
\begin{tabular}{lcccc}
\toprule
$\beta^2$ & $10^3\,[dE/dL-\eps(4\pi)]$ & $10^3\,\Delta\eps_0$ & $d\ell_0/dL$ & state\\
\midrule
\multicolumn{5}{l}{$a=1/2$}\\
$1.30$ & $-0.08\pm0.03$ & $+45.96$ & $0.00$ & bound\\
$1.58$ & $-0.15\pm0.06$ & $+2.91$ & $0.01$ & bound\\
$1.75$ & $-17.99$ & $-17.99$ & $1.00$ & stretched\\
$1.90$ & $-34.27$ & $-34.27$ & $1.00$ & stretched\\
\midrule
\multicolumn{5}{l}{$a=1/4$}\\
$1.30$ & $-0.08$ & $+46.98$ & $0.00$ & bound\\
$1.58$ & $-0.13$ & $+4.06$ & $0.00$ & bound\\
$1.75$ & $-16.76$ & $-16.76$ & $1.00$ & stretched\\
$1.90$ & $-32.96$ & $-32.96$ & $1.00$ & stretched\\
\bottomrule
\end{tabular}
\end{table}

\section{Discussion}
\label{sec:discussion}

Across the inversion line the $Q=n$ sector is a confinement problem with a string tension that changes sign. In the Ising field theory in a magnetic field~\cite{McCoy:1978,Fonseca:2003}, and in the perturbed sine-Gordon model where a $\cos(\beta\phi/2)$ term lifts the degeneracy of alternate vacua~\cite{Roy:2023,Rutkevich:2024}, kinks are confined into mesons by a positive string tension, whose value quantum fluctuations shift. At weak confinement the meson spectrum is set by the zeros of the Airy function, the first of which gives eq.~\eqref{eq:airy}. Here the tension of the central string is the sum of a classical part and a negative quantum part of comparable size. Below the inversion line the two halves of the multi-kink form such a meson, and above it they are no longer confined.

In the $\phi^6$ and $\phi^8$ models of refs.~\cite{Weigel:2017b,Takyi:2020}, and for the separated constituents of the Shifman--Voloshin kinks~\cite{AlonsoIzquierdo:2004,Weigel:2018}, the secondary vacuum inside the soliton is degenerate with the primary one and has a heavier meson. The quantum tension then acts alone, the one-loop energy has no lower bound at any coupling, and ref.~\cite{Takyi:2020} concludes that the soliton is destabilized at the quantum level. That case is the $\alpha\to0$ limit of the inversion line. The one-loop tension of eq.~\eqref{eq:eps1} evaluated between vacua of meson masses $1$ and $2$ is $-0.101$ per unit length, the rate at which the vacuum polarization energy of the $\phi^6$ kink changes with the position of its centre~\cite{Weigel:2017b}.

\section{Conclusions}
\label{sec:conclusions}

We have shown, beyond perturbation theory, that a composite soliton with a lifted interior vacuum loses its binding where the total tension of its interior string changes sign, while the correction to its mass stays small. The sign change is an inversion of the ordering of the two vacua. In the non-degenerate double sine-Gordon model one loop places it at a coupling of order $\alpha$. Computed with tensor networks for the same normal-ordered lattice Hamiltonian, the difference between the central and primary vacuum energy densities changes sign at a coupling below the one-loop value by a relative amount of order $\alpha$, and this crossing has a continuum limit. The Gaussian effective potential reproduces it to a fraction of a percent for every $n$ and $\alpha$ studied. On finite chains the tensor-network ground state shows the unbinding itself. As the tension falls the central string grows, and beyond the crossing the central vacuum fills the box and the energy changes with the length of the box at the rate set by the nonperturbative tension. At both bound couplings the mass of the multi-kink stays within ten percent of its classical value, so the mass does not signal the loss of binding and the sign of the tension does.

A similar competition arises whenever a soliton carries a region of lifted vacuum whose meson mass differs from the one outside. In the $\phi^6$ family of ref.~\cite{AlonsoIzquierdo:2011} a lattice calculation of the kind done here could test the inversion beyond one loop. Kink confinement has been studied in Ising chains, where a longitudinal field splits the vacua~\cite{Kormos:2017}, and observed in a trapped-ion simulator, where it comes from long-range interactions~\cite{Tan:2021}. Two vacua of unequal meson mass with a small splitting would put the inversion line within reach of the same bound-state spectroscopy. The multi-kinks of the classical model also bind into composite oscillons when they collide~\cite{LozanoMayo:2024dsg}. Whether a quantum counterpart survives its own radiation has been argued both ways, in each case at leading order in the coupling~\cite{Hertzberg:2010,Evslin:2023,Evslin:2025}. The lattice Hamiltonian used here can settle it nonperturbatively, by colliding two multi-kink wavepackets and following the collision region in real time as has been done for sine-Gordon solitons with tensor networks~\cite{Hechenberger:2026,Calliari:2025}.

\acknowledgments
I thank Manuel Torres-Labansat for introducing me to topological solitons and their quantum corrections.

\appendix

\section{One-loop energy density and the Gaussian effective potential}
\label{app:oneloop}

The zero-point energy density of a free field of mass $m$ with a momentum cutoff $\Lambda$ is $\mathcal{E}(m^2)=\frac12\int_{-\Lambda}^{\Lambda}\frac{dk}{2\pi}\sqrt{k^2+m^2}$. Its first derivative is $d\mathcal{E}/dm^2=\frac12G_\Lambda(m)$, with $G_\Lambda(m)=\int\frac{dk}{2\pi}\frac{1}{2\omega_k}$ and $\omega_k^2=k^2+m^2$. Its second derivative is $d^2\mathcal{E}/d(m^2)^2=-1/(8\pi m^2)+\ord(\Lambda^{-2})$, so $\mathcal{E}$ is concave in $m^2$. Normal ordering at $\mu$ subtracts $\frac12G_\Lambda(\mu)\,V''(\chi)$ from the potential, that is, the tangent of $\mathcal{E}$ at $\mu^2$. The one-loop density of a homogeneous field is therefore $\mathcal{E}(m^2(\chi))-\mathcal{E}(\mu^2)-\frac12G_\Lambda(\mu)[m^2(\chi)-\mu^2]$, which is finite as $\Lambda\to\infty$ and equals eq.~\eqref{eq:eps1}. Any other finite part of the counterterm adds a term linear in $m^2(\chi)$, which tilts the tangent and can change the ordering of two vacua with different $m^2$.

For the Gaussian effective potential, the expectation value of eq.~\eqref{eq:H} in a Gaussian state of mass $\Omega$ centred at $\bar\chi$ has the free part
\begin{equation*}
\begin{aligned}
\mathcal{E}(\Omega^2)-\tfrac12\Omega^2G_\Lambda(\Omega)
&-\big[\mathcal{E}(\mu^2)-\tfrac12\mu^2G_\Lambda(\mu)\big]\\
&=\frac{\Omega^2-\mu^2}{8\pi},
\end{aligned}
\end{equation*}
the expectation value of $\frac12\pi^2+\frac12(\partial_x\phi)^2$ in the Gaussian state of mass $\Omega$ minus its value in the free vacuum of mass $\mu$, which normal ordering sets to zero, and the cosine part is
\begin{equation*}
\begin{aligned}
\langle\Nmu\cos k\chi\rangle_\Omega
&=\cos(k\bar\chi)\exp\!\big[-\tfrac12k^2\beta^2\big(G_\Lambda(\Omega)-G_\Lambda(\mu)\big)\big]\\
&=\cos(k\bar\chi)\,(\Omega/\mu)^{k^2\beta^2/4\pi}.
\end{aligned}
\end{equation*}
Together they give eq.~\eqref{eq:gep}. The gap equation $\partial\eps_G/\partial\Omega^2=0$ reads
\begin{equation*}
  \Omega^2=-\sum_kc_kk^2(\Omega/\mu)^{k^2\beta^2/4\pi}\cos k\bar\chi .
\end{equation*}
At $\bar\chi=n\pi$ and $\mu=m_0$ it is solved by $\Omega=m_0$, at which $\eps_G=0$ and $\partial^2\eps_G/\partial(\Omega^2)^2=[m_0^2+(\beta^2/8\pi)\Vna''''(n\pi)]/(8\pi m_0^4)>0$. On the lattice, $\ln(\Omega^2/\mu^2)/4\pi$ in the exponent is replaced by $G_a(\mu)-G_a(\Omega)$ and $\mathcal{E}$ by the Brillouin-zone integral of $\frac12\omega_q$, with $G_a$ and $\omega_q$ of eq.~\eqref{eq:Ga}.

\section{Lattice Hamiltonian and tensor-network calculations}
\label{app:lattice}

On a chain of $N$ sites with spacing $a$ and $[\phi_j,\pi_l]=i\delta_{jl}/a$, the Hamiltonian is
\begin{equation}
  H=a\sum_{j=1}^{N}\left[\frac{\pi_j^2}{2}+\frac{(\phi_{j+1}-\phi_j)^2}{2a^2}+\frac{1}{\beta^2}\sum_k c_kZ_k\cos(k\beta\phi_j)\right],\qquad
  Z_k=\exp\!\left[\tfrac12k^2\beta^2G_a(\mu)\right],
  \label{eq:Hlat}
\end{equation}
with
\begin{equation}
  G_a(\mu)=\int_{-\pi/a}^{\pi/a}\frac{dq}{2\pi}\,\frac{1}{2\omega_q},\qquad \omega_q^2=\mu^2+\frac{4}{a^2}\sin^2\frac{qa}{2}.
  \label{eq:Ga}
\end{equation}
The factors $Z_k$ are the exact lattice normal-ordering constants, $\langle\cos k\beta\phi_j\rangle_\mu=Z_k^{-1}$ in the free lattice theory of mass $\mu$, and as $a\to0$, $G_a\simeq(1/2\pi)\ln(C/a\mu)$ and $Z_k\propto(a\mu)^{-k^2\beta^2/4\pi}$. In the rescaled variables $\chi_j=\beta\phi_j$, $p_j=\beta\pi_j$, $[\chi_j,p_l]=i(\beta^2/a)\delta_{jl}$,
\begin{equation}
  H=\sum_j\left[-\frac{\beta^2}{2a}\frac{\partial^2}{\partial\chi_j^2}+\frac{(\chi_{j+1}-\chi_j)^2}{2a\beta^2}+\frac{a}{\beta^2}V_Z(\chi_j)\right],\qquad
  V_Z=\sum_kc_kZ_k\cos k\chi,
  \label{eq:Hchi}
\end{equation}
which is the form implemented, with all energies those of the classical model. Each site carries the field on a sinc discrete-variable grid~\cite{Colbert:1992} of spacing $\Delta=0.45\,\beta$, on which every function of $\chi_j$ is diagonal, so the Hamiltonian is a matrix product operator of bond dimension three. The grid is centred on the middle of the field range, so it is mirror-symmetric about $\chi=0$ and the lattice Hamiltonian of the $Q=4$ sector keeps the reflection $\chi\to-\chi$, $x\to-x$. Table~\ref{tab:params} collects the parameters. Reducing $\Delta$ to $0.35\,\beta$ changes the energy of the bound state at $\beta^2=1.58$, $L=80$ by $4\times10^{-7}$. The string lengths change by $3$ and $5$ percent, almost entirely because the windows cover a whole number of grid points; for the central string this accounts for $0.34$ of the $0.35$ change.

\begin{table}[t]
\centering
\caption{Parameters of the tensor-network calculations. $\Delta$ is the spacing of the sinc grid, $d$ its number of points, and $D$ the bond dimension.}
\label{tab:params}
\small
\begin{tabular}{@{}l >{\raggedright\arraybackslash}p{0.31\textwidth} >{\raggedright\arraybackslash}p{0.46\textwidth}@{}}
\toprule
 & homogeneous vacua (VUMPS) & $Q=4$ sector (DMRG)\\
\midrule
chain & infinite, one-site unit cell & $N=L/a=120$ to $480$ sites, ghost sites at $\mp4\pi$\\
$\Delta$ & $0.45\,\beta$ & $0.45\,\beta$\\
$d$ & $33$ to $61$ & $66$ to $76$\\
$D$ & $8$ to $48$ or $64$; $8$ to $24$ for the central vacuum at $a=1/2$ & $8$, $16$, $32$, with checks at $48$\\
\bottomrule
\end{tabular}
\end{table}

The homogeneous vacua are computed with the variational uniform matrix product state algorithm~\cite{ZaunerStauber:2018} in MPSKit~\cite{MPSKit}, from a product of Gaussians centred on $\chi=n\pi$ or $\chi=0$. Uniform matrix product states were applied to the lattice $\phi^4$ theory in ref.~\cite{Milsted:2013}. The energy density of the central vacuum changes by less than $10^{-6}$ per site beyond bond dimension $D=24$. The primary vacuum is soft, with a correlation length of order the inverse meson mass $1/m_0$, and its energy is taken at the largest bond dimension. The last step of the bond-dimension ladder changes the energy by at most $3\times10^{-5}$ per site, which we take as the estimate of the remaining bond-dimension error. It moves the crossings by at most $4\times10^{-4}$.

The $Q=4$ sector is selected by ghost sites that hold $\chi=\mp4\pi$ beyond the two ends, and its ground-state energy and profile are obtained with single-site DMRG in MPSKit with subspace expansion. Each bond-dimension stage runs until the energy changes by less than $10^{-5}$ between sweeps and the string lengths by less than $0.002$ over two sweeps. At the two bound couplings the optimization is also started from halves pulled apart to $\ell_0=14$. All starts converge to the same state, with energies equal to within $10^{-5}$ and $\ell_0$ to within $0.02$. Raising the bond dimension from $32$ to $48$ at $\beta^2=1.58$ and $1.90$ changes the energies by at most $1.1\times10^{-3}$ and the string lengths by at most $0.014$.

\section{One-loop mass of the multi-kink}
\label{app:vpe}

The one-loop correction to the mass of $\Kn$ is the vacuum polarization energy of the fluctuation operator $-\partial_x^2+m_0^2+U(x)$, $U=\Vna''(\chi_K)-m_0^2$, in the no-tadpole scheme,
\begin{equation}
\begin{aligned}
  E^{(1)}[\Kn]&=\frac12\sum_j(\omega_j-m_0)-\int_0^\infty\frac{dk}{2\pi}\,\frac{k}{\omega_k}\left[\delta(k)-\delta^{(1)}(k)\right],\\
  \delta^{(1)}(k)&=-\frac{1}{2k}\int dx\,U(x),
\end{aligned}
  \label{eq:vpe}
\end{equation}
where $\delta$ is the sum of the even- and odd-channel phase shifts and $\omega_j$ are the bound-state frequencies, including the translational zero mode~\cite{Graham:2009,Weigel:2017}. The subtraction of the first Born approximation implements the counterterm, and in $1+1$ dimensions the corresponding Feynman diagram and the counterterm cancel identically in this scheme. In units of the meson mass $m$, the procedure gives $E^{(1)}=-0.33321\,m$ for the $\phi^4$ kink, against the exact $m(1/4\sqrt3-3/2\pi)=-0.33313\,m$, and $-0.31848\,m$ for the sine-Gordon kink, against the exact $-m/\pi=-0.31831\,m$.

For $n=4$ the fluctuation operator has a single bound state, the zero mode, and the $n-1$ relative sub-kink modes lie above the threshold $m_0^2$ as near-threshold resonances of the continuum. The correction grows in magnitude as $\alpha\to0$, as $\ln^2(1/\alpha)$ from the product of the string length and its zero-point density, but remains a small fraction of $M_{\rm cl}$ (table~\ref{tab:betac}). The string contribution, estimated as $\sum_i\ell_i\eps^{(1)}(\chi_i)$ with $\ell_i$ the length over which $|\chi_K-\chi_i|<1/2$, is $-0.34$ of the $-0.60$ at $\alpha=0.3$, and between one half and two thirds of $E^{(1)}$ over $0.05\le\alpha\le0.5$. The rest we attribute to the sub-kink cores, where $U<0$.

\bibliographystyle{JHEP}
\bibliography{references}

\end{document}